\documentclass[11pt]{article}
\usepackage{amsfonts}

\renewcommand{\theequation}{\arabic{section}.\arabic{equation}}

\def\be{\begin{equation}}
\def\ee{\end{equation}}
\def\bea{\begin{eqnarray}}
\def\eea{\end{eqnarray}}

\def\R{{\mathbb R}}

\def\Hlin{{\cal H}}
\def\Hproj{\overline{\Hlin}}
\def\coh#1{| #1 \rangle}
\def\sand#1#2#3{\langle #1 | #2 | #3 \rangle}
\def\paral{{|\!|}}
\def\B{{\bf B}}
\def\x{{\bf{x}}}
\def\Spin{{\bf S}}

\def\I{\hbox{I}}

\def\Hop{\widehat{H}}
\def\mycite#1#2{$\!{}^{#1}$}

\font\secfont=cmssbx10 scaled \magstep1
\begin{document}

\thispagestyle{empty}

\
%\hfill Sumbitted to JMP, 15 December 2000
%\hfill  Version 2.3, 15 December 2000. Sumbitted to JMP%\ \today
\hfill J. Math. Phys. {\bf 42}, No. 11, 5130 -- 5142 (2001)

\vspace{1cm}

\begin{center}
{\LARGE{\bf{Berry  Phase in homogeneous K\"ahler manifolds with
linear Hamiltonians }}}
\end{center}

\begin{center}
Luis J. Boya$^\dagger$, Askold M. Perelomov$^\ddagger$
\footnote{On leave of absence from the Institute for Theoretical
and Experimental Physics, 117259 Moscow, Russia}
and Mariano Santander$^\ddagger$
\end{center}

\begin{center}
{\it $^\dagger$ Departamento de F\'{\i}sica Te\'orica,
Facultad de Ciencias,\\  Universidad de Zaragoza,
E--50009 Zaragoza,  Spain}
\end{center}

\begin{center}
{\it $^{\ddagger}$ Departamento de F\'{\i}sica Te\'orica,
Facultad de Ciencias,\\ Universidad de Valladolid,
E--47011 Valladolid, Spain}
\end{center}

\vspace{0.5cm}
{emails: luisjo@posta.unizar.es, perelomo@dftuz.unizar.es,
santander@fta.uva.es}

\vspace{0.5cm}

\bigskip\bigskip

\begin{abstract}\noindent
We study the total (dynamical plus geometrical (Berry)) phase of
cyclic quantum motion for coherent states over homogeneous
K\"ahler manifolds
$X=G/H$, which can be considered as the phase spaces of classical
systems and which are, in particular cases, coadjoint orbits of
some Lie groups $G$. When the Hamiltonian is linear in the
generators of a Lie group, both phases can be  calculated exactly
in terms of {\em classical} objects. In particular, the geometric
phase is given by the symplectic area enclosed by the (purely
classical) motion in the space of coherent states.

\end{abstract}

\bigskip\bigskip

%\newpage

%%%%%%%%%%%%%%%%% Introduction %%%%%%%%%%%%%%%%%%%%%%%%%%%%

\section*{\secfont I. INTRODUCTION}

Let us consider a quantum state ${\psi(t)}$ whose evolution
follows a time--dependent Schr\"odin\-ger equation. If the final
state $\psi (T)$ coincides with initial one $\psi (0)$, then  the
representative state vectors $\coh{\psi (0)}$ and $\coh{\psi
(T)}$ differ one from another just by a  phase factor
$\exp (i\alpha )$. This phase factor can be splitted into two
parts
$\alpha =\beta +\gamma $,  called respectively dynamical phase
and geometrical phase. Both $\beta$ and $\gamma$ are important
characteristics of the evolution of the system under
consideration.

In particular the geometric phase turns out to depend on the
Hamiltonian in a rather indirect way, as it is determined {\em
only} by the closed loop traversed by the state in the state
space.   This geometrical phase associated to any quantum cyclic
motion with time-dependent Hamiltonians appears, in addition to
the well-known dynamical phase, due to the natural curvature of
the line bundle over the projective Hilbert space of states. This
was found by Berry
\mycite{1}{Be} for adiabatic motion, interpreted by Simon
\mycite{2}{Simon}
as above, and extended by Aharonov-Anandan in \mycite{3}{AA} (see also
\mycite{4,5}{BoBoKe,GPP} for arbitrary cyclic motion). However, there
are very few cases in which the calculation can be performed
explicitly, and it would be nice to exhibit examples where the
phases of a cyclic quantum motion can be calculated in closed
terms.

We shall consider the important cases in which  the Hamiltonian
$H(t)$ is {\em linear} in the generators of a Lie algebra  $\cal
G$ acting  through  some unitary irreducible representation
$T^\lambda $  in a Hilbert space ${\cal H}^\lambda $, where
$\lambda $ labels the  representation. The aim of this paper is
to show that in these cases, explicit expressions for both
$\beta $  and $\gamma $ can be given in terms of a related {\em
classical} dynamical system. This is achieved by using the
generalized coherent state technique \mycite{6,7,8}{Pe1, Pe2, Pe3}, and is
done in a frame general enough to cover a wide variety of
examples and particular cases, including the well-known situation
for evolution of a spin 1/2 in a magnetic field, an standard
example which is however an oversimplified one, because its
quantum state space is the Riemann sphere $\mathbb C P^1$.

Therefore all information  on {\em dynamical and geometrical
phases} for these quantum systems   can be obtained by studying
the motion of a  {\it purely  classical} system in a suitable
phase space. As we shall see these  are the K\"ahler (and hence
naturally symplectic) homogeneous spaces $X = G/H$, with $G$  the
Lie group of the Lie algebra  $\cal G$. Important examples of such
spaces are the orbits of the coadjoint representation of compact
semisimple Lie groups. For
$G=U(n)$ the generic (maximal dimension) coadjoint orbit is
$U(n)/ U(1)^n$; this space is called a flag manifold, and plays
an important role in many areas
\mycite{9,10}{Bo2, CdWitt}.

The set-up of this paper is as follows: in Section 2 we present
the main ideas leading to closed expressions for dynamical and
geometrical phases, in terms of motion in the space $X$ taken as
a {\em classical} space. This is possible  when the quantum
Hamiltonian is linear in the generators of some representation of
a Lie algebra $\cal G$ and besides $X$ is an homogeneous K\"ahler
manifold of the Lie group $G$. In Section 3 we describe some
homogeneous  K\"ahler manifolds; they include: (i) coadjoint
orbits of semisimple compact Lie groups, (ii) the so-called
bounded symmetric domains which are not compact, and (iii) some
other cases, like the Heisenberg ``plane".

Finally in the Appendix A we collect explicit expressions for the
kernels which determine the K\"ahler potential, and we give some
differential and topological information on K\"ahler manifolds,
including the Poincar\'e polynomials. A resum\'e of relevant
details on coherent states, extracted from \mycite{8}{Pe3}, is also
included as Appendix B.

\section*{\secfont II. THE GROUP THEORETICAL COMPUTATION OF PHASES}
\setcounter{equation}{0}
\renewcommand{\theequation}{2.\arabic{equation}}

Let us consider the time--dependent Schr\"odinger equation
\be \label{s1}
 i\,\frac{d}{dt}\,|\psi (t)\rangle =\Hop(t)\,|\psi (t)\rangle,
\ee
with a Hamiltonian of the form:
\be
\Hop(t)= \sum_j a_j(t)\,X_j^\lambda ,\qquad
X_j^\lambda = {\cal T}^\lambda(X_j),
\ee
where ${\cal T}^\lambda $ is an unitary irreducible
representation of the Lie  algebra $\cal G$, whose generators
$X_j$ are represented in ${\cal T}^\lambda $ by the
(antihermitian) operators $i X_j^\lambda$ and $a_j(t)$ are arbitrary real
functions
of time. We consider here only those cases when the representation
Hilbert space
${\cal H}^\lambda $ may be realized as a space ${\cal F}^\lambda
$ of  holomorphic functions on a complex homogeneous space
$X=G/H$ which is  also a K\"ahler one. We assume also that the
initial state is a  generalized coherent state $\coh{x_0}$
labeled by the point $x_0\in X$; for details, see \mycite{8}{Pe3}.

In this case, under time evolution the initial coherent state
remains a coherent state
\be
|x(t)\rangle = U(t, 0) \,|x(0)\rangle
\ee
and then $x(t)$ is a solution of the Hamilton equation for the
corresponding  {\em classical\/} system
\be \dot x=\{ H(t),x\} ^\lambda,\qquad \dot x =\frac{dx}{dt},
\label{clas}
\ee
where $\{\,,\,\}^\lambda $ is the Poisson bracket induced on $X$
by the  representation ${\cal T}^\lambda $.

The mapping $X \to \Hlin^\lambda $ which associates the point
$x_0 \in X$ to the coherent state $\coh{x_0}$ allows an isomorphic
identification of actual quantum ``trajectories" starting from
$\coh{x_0}$ and obeying the usual Schr\"odinger equation
\be
 i \frac{d}{dt} \coh{x(t)} = \Hop(t)\coh{x(t)}
\ee
to some {\em classical\/} motions in $X$ (taken as a {\em
classical\/} phase space, not a configuration space), satisfying
(\ref{clas}).

Under this identification, if $\Gamma$ is a closed loop in $X$ with
period $T$,
it is still closed in the projective Hilbert space
$\Hproj^\lambda $, which should be considered as the {\em true}
state space,  but not necessarily in the linear Hilbert space
$\Hlin^\lambda $. In this cyclic motion, the state {\em vector}
picks up a phase
\be
        \coh{\psi(T)} = \exp{(i \alpha)} \coh{\psi(0)}, \qquad
\alpha=\beta+\gamma.
\ee
This can be also seen as follows \mycite{4}{BoBoKe}.
Let $\Gamma$ be a closed path (loop) in the projective Hilbert
space
$\Hproj = {\mathbb C}P^\infty$ of states; let $\coh{\psi}
=
\coh{\psi(t)}$ be a generic point in $\Gamma$. There is a
tautological line bundle, whereby each point carries its vectors;
this line bundle is hermitian, by the hermitian product in
$\Hlin $. Let $P(t)$ be in the fibre over
$\coh{\psi(t)}$. The Hamiltonian $\Hop$ works in
$\Hlin $, and by projection in $\Hproj $   also,
so the time evolution carries $P(0) \to P(t)$ and projects to
$U(0, t):
\coh{\psi(0)} \to \coh{\psi(t)}$. As $P(T)$ is the fibre
over
${\psi(T)}$ which coincides with the fibre over ${\psi(0)}$, we
must have:
\be
            P(T) = \exp{(i\alpha)} P(0) ,
\ee
where $\alpha$ is the total phase for the cyclic motion. The lift
of the path $\Gamma$ through the connection of the line bundle
$L^\lambda$ would produce an $U(1)$ holonomy $\gamma$;  this is
the {\em geometric phase}, and the difference, $\beta = \alpha -
\gamma$ is the dynamical phase. As explained in detail in
\mycite{2}{Simon}, we have the following explicit expressions for both
parts of the total phase $\alpha$:
\be
\beta = \int \sand{\psi(t)}{\Hop(t)}{\psi(t)} dt,   \qquad
\gamma = \int \sand{\psi'(t)}{\Big( -i\,\frac{d}{dt}\,\Big) }{\psi'(t)}
dt,
\ee
where $\coh{\psi'(t)}$  is a trivializing section, i.e. there is
no dynamical phase for the whole loop, see \mycite{4,11}{BoBoKe,BoCaGra}.
The
connection 1-form of this line bundle $\theta^\lambda$ is related
locally to the symplectic 2-form as
$\omega^\lambda=d
\theta^\lambda$, and this symplectic 2-form is in turn induced by
the imaginary part of the Hermitian scalar product in
$\Hlin^\lambda $.

Under the conditions stated, both phases can be computed directly
in terms of the classical motion in $X$. For the dynamical part
we have:
\be
\beta = \int \langle \psi (t)|\,\Hop(t)\,|\psi (t)\rangle \,dt\\
      = \int \langle x(t)\,|\Hop(t)\,|x(t)\rangle \,dt\\
         = \int a_j(t)\,X_j^\lambda (t)\,dt,
\ee
where $X_j^\lambda (t)=\langle x(t)|\,X_j^\lambda \,|x(t)\rangle
$.

The geometric phase $\gamma$ is given as the integral along
$\Gamma$ of the connection 1-form $\theta^\lambda$ which depends on
the representation. Due to the abelian nature of the $U(1)$
group, the Stokes theorem applies and gives:
\be
\gamma =\gamma _{cl}=\int _\Gamma \theta^\lambda =\int _{\Sigma }
\omega ^\lambda ,
\label{Gphase}
\ee
where $\omega ^\lambda = d \theta^\lambda$ and $\Sigma$ is any
surface having $\Gamma =\partial \Sigma $ as its  boundary.
Hence, we give the expression for the geometric phase in terms
of  symplectic area of any surface whose boundary is the given
(classical)  closed circuit in our K\"ahler manifold $X$.

Thus formula (\ref{Gphase}) is valid for arbitrary
homogeneous K\"ahler manifolds.  Now if $F^\lambda$ denotes the
K\"ahler potential \mycite{12,13}{Chern, Nak}, the expressions for
the connection and curvature forms are (see Appendix A):
\be
\theta^\lambda =\frac{1}{i}\left(
\frac{\partial F^\lambda}{\partial z_\mu }\,dz_\mu -\,
\frac{\partial F^\lambda}{\partial \overline z_\mu }\,
d\overline z_\mu \right)
,\qquad
\omega^\lambda =\frac{i}{2}
\frac{\partial ^2F^\lambda}{\partial z_\mu
\,\partial \overline z_\nu }
\,dz_\mu \land \partial\overline z_\nu =d\theta^\lambda,
\ee
and the K\"ahler potential itself is related to the kernel,
which generalizes Bergmann's kernel,  as:
\be
F(z, \overline z) =
\left. \ln \left( K(z, \overline w) \right) \right|_{w=z}.
\ee
The simplest closed loops are geodesic triangles. For them we can
give explicit expressions.

Let us first consider the simplest case ${\cal G}=su(2)$, where
$X=\mathbb C P^1=SU(2)/U(1)$ is the Riemann sphere, and
$U(1) \to SU(2)=S^3 \to S^2$ is
the second Hopf sphere bundle.
In this case
there is a single complex coordinate $z$, related with the point
$x$ on the sphere by the usual stereographic projection, and the
Bergmann's kernel is  given by
\be
K(z, \overline w) = 1 + z \overline w.
\ee

It is clear that any vertex can be carried to a prescribed point
on the sphere, say the North pole, corresponding to $z=0$. Let
$\coh{x}, \coh{y}$ denote the two coherent states determining the
remaining triangle vertices, corresponding to points $x, y$ on
the sphere, and let us denote $z, w$ the complex coordinates
corresponding to $x, y$. The closed expression for the geometric
phase associated to this closed loop is:
\be
\gamma = \frac{1}{2i} \ln \frac{K(\overline z, w)}{K(z, \overline w)} =
\frac{1}{2i} \ln \frac{1+\overline z w }{1+z \overline w} =
\frac{1}{2i} \ln \frac{\langle x | y \rangle}{\langle y | x \rangle}.
\ee

For ${\cal G}=su(2)$ this  result for the phase has been given
already in \mycite{6}{Pe1} (see also  \mycite{8}{Pe3}). In this case the
symplectic area is proportional to the riemannian area for the
standard riemannian structure on $S^2$; this proportionality is
however accidental and may be misleading because this does not
hold in higher dimensions; for instance in $\mathbb C P^n$ $(n>1)$ the
symplectic area of any finite triangle is not proportional to its
riemannian
Fubini-Study area.

Let us now consider the general case where the generators
$X_j^\lambda$ close to an unitary irreducible representation of the
Lie  algebra $\cal G$. The symplectic area of any closed loop in
$\Hproj$ is completely determined by the loop; this does not hold
for the `riemannian' area determined by a general K\"ahler
metric, which depends essentially on the two-dimensional surface
whose boundary is the prescribed loop. In this general case
(with group $\cal G$), it suffices again to give a closed
expression for the symplectic area of  a triangular loop. If one
vertex is carried to a prescribed point on the K\"ahler
homogeneous manifold  $X$ (say determined by the complex
coordinates $z_\mu=0$), the remaining vertices
$x, y$ will correspond to the complex coordinates
$z_\mu, w_\nu$. The same argument as before leads in this case to
the expression:
\be
\gamma \propto  \frac{1}{2i} \ln
\frac{K(\overline z, w)}{K(z, \overline w)}.
\label{GeomPhase}
\ee

Appendix A contain explicit expressions for the kernels $K(z,
\overline w)$ for Hermitian symmetric spaces, taken from
\mycite{8}{Pe3}, where further details on the construction of the
kernel $K(z_\mu, \overline w_\nu)$ for homogeneous K\"ahler
manifolds can be found.

In the case of complex grassmannians, the usual choice for
complex coordinates are called  Pontrjagin coordinates and can be
arranged as a complex rectangular matrix $Z$. After substituting
for the relevant kernel, the basic expression
(\ref{GeomPhase})  reduces to:
\be
\gamma \propto
\frac{1}{2i} \ln
\frac{\det( \I + Z W^\dagger)}{\det( \I + W Z^\dagger)}
\ee
and coincides with the formula for the geometric phase derived
through explicit computation  by Berceanu, who also points out
the  validity of a similar formula for  any Hermitian symmetric
space \mycite{14}{Be1}.  However, the  arguments given in \mycite{8}{Pe3}
appear to hold unrestrictedly for arbitrary  homogeneous
K\"ahler manifold $X$, and not only for Hermitian symmetric
spaces.

In the well understood example (see e.g. \mycite{15, 16}{FeNiOlSa,
FeOlSa}), of a spin
$1/2$ particle in a magnetic field,
\be
        i \, \frac{d}{dt}\, \coh{\psi(t)} = - \mu \, \B(t) \cdot \Spin
\coh{\psi(t)}
\ee
the Hamiltonian is a linear combination of three operators which
span a Lie algebra $su(2)$, and quantum evolution can be thus
translated into a classical motion of a point on  the homogeneous
space $SU(2)/U(1)$, the Riemann sphere again. The coadjoint orbits are
2-spheres and $x = \x$ is a unit vector in $\R^3$, so at any fixed
time $\Hop(t)$ splits into two parts:
\be
\Hop(t) = \Hop_\paral  +  \Hop_\perp,
\ee
where
\be
\Hop_\perp \, \coh{\x(t)} = 0, \qquad
\Hop_\paral\, \coh{\x(t)} = E(t)
\, \coh{\x(t)}.
\ee

The longitudinal part produces only a dynamical phase, as the ray
of $\coh{\x}$ and of $E(t)\coh{\x}$ are the same. The geometrical
phase comes entirely from the transverse part.  In particular, if
the field is constant in direction:
\be
        H(t) = B(t) \, \sigma_z
\ee
and the initial state is $\coh{\x}= \cos(\theta/2) \, \coh{+} +
\sin(\theta/2) \, \coh{-}$, the solution is readily obtained
\be
\coh{\psi(t)}= a(t) \, \coh{+} + b(t) \, \coh{-},
\ee
where $a(t) = a(0) \exp(-i\int B(t) \, dt),\    b(t) = b(0) \exp(i
\int B(t)\,  dt)$. For $\theta=0$ or $\theta=\pi$ we have a
purely dynamical phase, while for $\theta=\pi/2$ the phase is
purely geometrical.

For arbitrary $\B(t)$ there is also a local splitting, and the
``parallel" $H_\paral$ and ``perpendicular" $H_\perp$ parts  of
the Hamiltonian carry respectively the dynamical and  geometric
phases.

\section*{\secfont III HOMOGENEOUS SYMPLECTIC MANIFOLDS AND K\"AHLER
MANIFOLDS}
\setcounter{equation}{0}
\renewcommand{\theequation}{3.\arabic{equation}}

Relative to the definition of a symplectic manifold, see the book
\mycite{17}{Ar}.

{\bf Definition}. {\em A symplectic manifold $(M,\omega )$  is
called homogeneous  if there exists on it a transitive action
$\Phi _g:M\to M$ of some  Lie group $G=\{g\}$ which acts as a
group of symplectic transformations,  i.e., it leaves invariant
the form $\omega $, $\Phi _g^*\,\omega =\omega $.}

{\bf Theorem} \mycite{18}{Ki}. {\em Any homogeneous symplectic manifold
on which  a  connected Lie group $G$ acts transitively and by
symplectic transformations is locally isomorphic to an  orbit of
a coadjoint representation of this group
$G$ or of a central extension of $G$ by ${\mathbb R}$}.

Thus any coadjoint orbit of the group $G$ is an homogeneous
symplectic manifold.

Among the class of all homogeneous symplectic manifolds, the main
important subclass is those of coadjoint orbits of {\em
semi-simple} Lie groups. These have an additional K\"ahler
homogeneous structure. A K\"ahler manifold is  defined as a
complex manifold $M$ endowed with a K\"ahler metric $h$, whose
imaginary part is a closed two-form. A K\"ahler metric is an
hermitian metric $h$  which comes from a function $F(z,{\overline
z})$ called the K\"ahler potential:
\be
ds^2=h_{\mu \overline \nu } \,dz^\mu \,d\overline {z}^\nu ,\quad
h_{\mu \overline \nu } (z,\overline {z})=\partial _\mu \,
\partial_{\overline \nu }\,F(z,{\overline z}),\quad \partial _\mu =
\frac{\partial }{\partial z_\mu },\quad \partial _{\overline \nu }=
\frac{\partial }{\partial {\overline z}_{\nu }}.
\label{metric}
\ee

The imaginary part of this metric is a symplectic 2-form
\be
\omega =\frac{i}2\,
h_{\mu \overline {\nu }}(z,{\overline z})\,dz^\mu \land d{\overline
z}^\nu ,
\qquad
d\omega=0.
\label{twoform}
\ee

The connection between orbits of the coadjoint representation of
compact simple Lie groups and K\"ahler homogeneous manifolds is
stated in the following important result of A. Borel:

{\bf Theorem} \mycite{19}{Bo1}. {\em Any orbit of the coadjoint
representation of a  compact simple Lie group  is a compact
K\"ahler homogeneous simply-connected manifold,  and any compact
K\"ahler homogeneous simply-connected manifold is some orbit  of
the coadjoint representation of the some compact simple Lie
group}.

Orbits of the coadjoint representation of a compact Lie group are
even rational manifolds \mycite{20}{Go}. Topologically they are compact
and simply-connected manifolds. Their topology  is described, for
example, in the review \mycite{9}{Bo2}. In the Appendix A we give some
pertinent results.

Many examples of K\"ahler homogeneous manifolds with a compact
group $G$ are known; these spaces are compact, even-dimensional,
simply-connected and oriented. As the cohomology class $[\omega]
\neq 0$, all the even Betti numbers are nonzero. Let us recall
some simple examples.

For $G = SU(2) = Spin(3) \sim SO(3)$, the generic coadjoint
orbits in $su(2) \approx \R^3$ are spheres $S^2$; there is an
isolated orbit consisting of a single point, the origin. For each
sphere the 2-form is just the area (volume)  form, automatically
closed by dimensionality. It is a complex (one-dimensional)
manifold, the Riemann sphere.

For $G = SU(3)$, there are three types of coadjoint orbits in
$su(3) \approx \R^8$: the origin, four-dimensional orbits
isomorphic to $\mathbb C P^2 = \frac{SU(3)}{SU(2)\times U(1)}$,
and six-dimensional maximal orbits, isomorphic to the flag
manifold ${\mathbb F}^3=\frac{SU(3)}{U(1)\times U(1)}$.

For $G=SU(n)$, the description of the orbits is essentially
given by the partitions of $n$ (see \mycite{21}{AP}).

The general calculation of K\"ahler metrics on the coadjoint
orbits for any compact simple Lie group (the classical and
exceptional structures of Cartan) was expelled in \mycite{21}{AP}.

The main reason why these manifolds are K\"ahler is that the
homogeneous structure is also obtained from the complex extension
$G^{\mathbb C}$ of $G$. The role of the subgroup $H$ here is played by
some
triangular (Borel) subgroup $B$; both $G^{\mathbb C}$ and $B$ are
analytic manifolds, and so is $G^{\mathbb C}/B$ which turns out
to be isomorphic to $G/H$. The space $X$ is also obviously simply
connected, because $G$ can be taken simply connected (for any
$X$) and $H$ is connected.  This construction $X=G/H = G^{\mathbb
C}/B$ is also basic in the Borel-Weil-Bott theory of analytic
construction of irreducible representations of $G$ as sections in
some holomorphic bundles.

When $G$ is a general simple or semisimple compact Lie group, the
orbits of the coadjoint representation  exahusts all the compact
homogeneous K\"ahler manifolds.

Other examples of (non-compact) K\"ahler manifolds are the
so-called bounded symmetric domains (see \mycite{22}{He}). Recall that
a bounded domain $D \subset \mathbb C^n$ is called symmetric if
each point in D is fixed by an involutive holomorphic
diffeomorphism of D. These are characterized by the result:

{\bf Theorem} [Helgason \mycite{22}{He}, p.310]. {\em  (i) Each
bounded symmetric domain $D$, when equipped with the Bergmann
metric, is a Hermitian symmetric space of the non-compact type. In
particular, a bounded symmetric domain is neccesarily simply
connected.

(ii) Let $M$ be a Hermitian symmetric space of the non-compact
type. Then there exists a bounded symmetric domain $D$ and a
holomorphic diffeomorphism of $M$   onto $D$}.

The paradigmatic example is the Lobachevsky plane. This a
K\"ahler  manifold which is non-compact, and of constant negative
curvature.

A complete classification of K\"ahler manifolds is still lacking.
Hermitian symmetric spaces, which are completely classified, are
examples of K\"ahler manifolds, while the remaining non-hermitian
symmetric spaces are not K\"ahler (e.g, the even dimensional
spheres $S^{2n}, \, n>1$ are homogeneous and symmetric, but
obviously not K\"ahler).

Some nonsemisimple groups also provide other K\"ahler manifolds.
A very basic example is that obtained from the
Heisenberg-Weyl algebra $hw(1)$ generated by the usual operators
$p, q, 1$, by quotient by the subgroup generated by the
subalgebra $1$. This space is the basic ``quantum" space $q, p$,
whose non-compact  K\"ahler character becomes obvious after
introduction of the complex coordinate
$z=p+i q$.

\setcounter{equation}{0}
\renewcommand{\theequation}{A.\arabic{equation}}

\section*{\secfont APPENDIX A. K\"AHLER HOMOGENEOUS MANIFOLDS}

We start by listing some examples of compact K\"ahler homogeneous
manifolds. More details can be found in \mycite{19}{Bo1} and in
\mycite{21}{AP}.

\begin{enumerate}

\item $G=SO(3)\sim SU(2)$ is the rotation group of a
three-dimensional  vector space ${\mathbb R}^3$. Here the sign
$\sim$ means a locally isomorphic group and ${\cal G}^*$ is the
dual space to the Lie algebra
${\cal G^*}=\{  {\bf x}\vert {\bf x}=(x_1,x_2,x_3)\} \approx
\R^3$. There is a zero-dimensional orbit (the origin) while the
remaining orbits are generic and are  two-dimensional spheres
$S_r^2=\{ {\bf x} \vert {\bf x}^2= x_1^2+x_2^2+x_3^2 =r^2 \}$.

\item $G=SU(3)$. Here we have three types of coadjoint orbits in
$su(3) \approx \R^8$: First, the origin ${\bf x}=0$. Second,
four-dimensional orbits  (isomorphic to ${\mathbb C}P^2$)
\be
{\cal O}=\frac{SU(3)}{SU(2)\times U(1)}\,,
\ee
and third, six-dimensional orbits isomorphic to the complex flag
space ${\mathbb F}^3$
\be
{\cal O}=\frac{SU(3)}{U(1)\times U(1)}\,.
\ee

\item $G=SU(n)$. Here, in addition to the trivial
zero-dimensional orbit, we have orbits isomorphic to the  complex
projective space ${\mathbb C}P^{n-1}$,
\be
{\cal O} =\frac{SU(n)}{SU(n-1)\times U(1)}\sim {\mathbb C}\,P^{n-1}\,.
\ee
There are also orbits isomorphic to the complex grassmannians
$\mathbb C G_{m,n}$,
\be
{\cal O}=\frac{SU(m+n)}{SU(m)\times SU(n)\times U(1)}\sim \mathbb
C G_{m,n},
\ee
and finally the generic maximal orbits are isomorphic to the
complex flag manifold ${\mathbb F}^{n}$
\be
{\cal O}=\frac{SU(n)}{U(1)\times U(1)\times\cdots
\times U(1)}\sim {\mathbb F}^{n}.
\ee

\item For compact simple Lie algebras, the coadjoint orbits of
minimal non-zero dimension were investigated in  \mycite{23}{Wo}, and
are given in the following Table:

\medskip

\begin{tabular}{ll|l|l}
$G$&& $\quad \mbox{dim}\,{\cal O}_{\mbox{min}}$& $\quad H$ \\[5pt]
\hline\hline \\[-10pt]
$A_n$&$\quad SU(n+1)\quad $ &$\quad 2n\quad $&$\quad A_{n-1}\times U(1)
\quad $\\
$B_n$&$\quad SO(2n+1)\quad $&$\quad 2(2n-1)\quad $&$\quad B_{n-1}\times
SO(2)\quad $\\
$C_n$&$\quad Sp(n)\quad $&$\quad 2(2n-2)\quad $&$\quad C_{n-1}\times
U(1)\quad $\\
$D_n,\,\,n\neq 2$&$\quad SO(2n)\quad $&$\quad 2(2n-2)\quad $&$\quad
D_{n-1}\times
SO(2)\quad $ \\
$G_2$&$\quad $&$ \quad 10\quad $&$\quad A_1\times SO(2)\quad$ \\
$F_4$&$\quad $&$\quad 30\quad $&$\quad C_3\times SO(2)\quad $ \\
$E_6$&$\quad $&$\quad 32\quad $&$\quad D_5\times SO(2)\quad $\\
$E_7$&$\quad $&$\quad 54\quad $&$\quad E_6\times SO(2)\quad $\\
$E_8$&$\quad $&$\quad 114\quad $&$\quad E_7\times SO(2)\quad $\\[5pt]
\hline\hline \\
\end{tabular}

\end{enumerate}

\subsection*{A.1 Kernels for some Hermitian symmetric spaces}

In this section we give the explicit expressions \mycite{24}{Hu} for
kernels of the Hermitian symmetric spaces of classical type,
either compact or non-compact (bounded symmetric domains). They
belong to four families, which in the Cartan notation are
$A_{III}, C_{I}, D_{III}$ and $BD_{I}(q=2)$ \mycite{22}{He}. There are two
further exceptional Hermitian symmetric spaces, $E_{III},
E_{VII}$ related to exceptional Lie algebras.

\begin{description}

\item[$A_{III}$c]\ \par For the complex grasmannians $\mathbb C G_{p,q}$
of $p$-planes in $\mathbb C^{p+q}$:
\be
X=SU(p+q)/(SU(p) \otimes SU(q) \otimes U(1)), \quad p\geq q,
\ee
in terms of the $pq$ complex coordinates arranged in a
rectangular $p\times q$ complex matrix $Z$:
\be
K(z, \overline w) = \det( \I^{(p)} + Z W^\dagger).
\ee

\item[$A_{III}$nc]\ \par  The  non-compact Cartan duals of the complex
grassmannians are the spaces:
\be
X=SU(p,q)/(SU(p) \otimes SU(q) \otimes U(1)), \quad p\geq q.
\ee
which can be realized as the bounded domain $\I^{(p)} - Z
Z^\dagger \geq 0$ with $Z$ as above; its kernel is:
\be
K(z, \overline w) = \det( \I^{(p)} - Z W^\dagger).
\ee

\item[$C_{I}$c]\ \par  For the manifold of
Lagrangian $p$-spaces in $\mathbb C^{2p}$, which is the compact
symmetric Hermitian space:
\be
X=Sp(p)/U(p)
\ee
the kernel is given in terms of $p(p+1)/2$ complex coordinates
arranged in a  $p\times p$ complex symmetric matrix $Z$ as:
\be
K(z, \overline w) = \det( \I^{(p)} + Z W^\dagger).
\ee

\item[$C_{I}$nc]\ \par  The Cartan dual to the previous space:
\be
X=Sp(2p, \mathbb R)/U(p)
\ee
can be realized as the bounded domain $\I^{(p)} - Z Z^\dagger
\geq 0$ in terms of the coordinate matrix  $Z$ as above; its
kernel is:
\be
K(z, \overline w) = \det( \I^{(p)} - Z W^\dagger).
\ee

\item[$D_{III}$c]\ \par  The kernel for the compact Hermitian symmetric
space:
\be
X=SO(2p)/U(p)
\ee
is given in terms of $p(p-1)/2$ complex coordinates arranged in a
rectangular $p\times p$ complex skew-symmetric matrix $Z$ as:
\be
K(z, \overline w) = \det( \I^{(p)} + Z W^\dagger).
\ee

\item[$D_{III}$nc]\ \par  For the non-compact Cartan dual space:
\be
X=SO^*(2p)/U(p)
\ee
realized as the bounded domain $\I^{(p)} - Z Z^\dagger \geq 0$ in
terms of the coordinates $Z$ as above, the kernel is:
\be
K(z, \overline w) = \det( \I^{(p)} - Z W^\dagger).
\ee

\item[$BD_{I}$c]\ \par  The real grasmannian $\mathbb R G_{2,p}$ of
2-planes in $\mathbb R^{p+2}$:
\be
X=SO(p+2)/(SO(p)\otimes SO(2))
\ee
In terms of $p$ complex coordinates arranged as a $p\times 1$ row
complex vector $\bf{z}$, with
$\bf{z}'$ denoting the transpose $1\times p$ column complex
vector, then
\be
K(z, \overline w) = 1 + ({\bf{z}} \cdot {\bf{z}}')
(\overline{\bf{w}} \cdot \overline{\bf{w}}') +
2\, ({\bf{z}} \cdot \overline{\bf{w}}').
\ee

\item[$BD_{I}$nc]\ \par  The non-compact dual space:
\be
X=SO(p, 2)/(SO(p)\otimes SO(2))
\ee
can be realized as the bounded domain
\be
\mid {\bf{z}} \cdot {\bf{z}}' \mid <1, \qquad
1+ \mid {\bf{z}} \cdot {\bf{z}}' \mid^2 -
2 \, \overline{\bf{z}} \cdot {\bf{z}}' >0,
\ee
where the $p$ complex coordinates are arranged as a $p\times 1$
row complex vector $\bf{z}$, as above; the kernel is:
\be
K(z, \overline w) = 1 + ({\bf{z}} \cdot {\bf{z}}')
(\overline{\bf{w}} \cdot \overline{\bf{w}}') -
2(\bf{z} \cdot \overline{\bf{w}}').
\ee

\end{description}

The two exceptional Hermitian symmetric spaces can be dealt with
similarly, by using $3 \times 3$ octonionic matrices, as
discussed by U. Hirzebruch \mycite{25}{Hirz}.

\subsection*{A.2 Topology of orbits}

Orbits of a coadjoint representation of compact Lie groups are
compact simply-connected manifolds; this follows from the exact
homotopy sequence. They have a non-trivial second homotopy group
$\pi_2(X)$
because they are compact symplectic manifolds. Further
information on their topology  may be found, for example, in the
review \mycite{9}{Bo2}.

Let $P_X(t)=\sum _{j=0}^N b_j\,t^j$ be the Poincar\'e polynomial
of manifold  $X$, $b_j$ being the Betti numbers of the manifold
$X$ of dimension $N$. In our case $X=G/ H$, where $H$ is some
compact semisimple subgroup of $G$, and
$\mbox{rank}\,H=\mbox{rank}\,G=r$. In this case, the Hirsch
formula (see \mycite{9}{Bo2}) is valid
\be
P_X(t)=\frac{\prod _{j=1}^r\,\left( 1-t^{2n_j}\right) }{\prod
_{j=1}^r\,
\left( 1-t^{2m_j}\right) }\,,
\ee
where $n_j$ and $m_j$ are the degrees of basic invariants of the
Weyl group $W$ of  the groups
$G$ and $H$ (see \mycite{26}{Ch}). Let us give a few applications of
this formula. We have

\begin{description}

\item[(i)] For the complex projective space:
\be
X={\mathbb C}\,P^n, \quad
P_X(t)=P_n(t)\equiv 1+t^2+t^4+\cdots
+t^{2n}.
\ee

\item[(ii)] For the complex flag manifold ${\mathbb F}^n$:
\be
X={\mathbb F}^n=\frac{SU(n)}{U(1)\times \cdots \times
U(1)},\qquad  P_X(t)=P_1(t)\,P_2(t)\,\cdots \, P_{n-1}(t),
\ee
where the polynomial $P_n(t)$ was defined above.

\item[(iii)] An example of a real grassmannian $\mathbb R
G_{3,2}$:
\be
X=\frac{SO(5)}{SO(3)\times SO(2)}, \qquad
P_X(t)=P_3(t).
\ee

\item[(iv)] An example of a real ``flag-like" manifold:
\be
X=\frac{SO(5)}{SO(2)\times SO(2)},\quad P_X=P_1(t)\,P_3(t).
\ee

\item[(v)] For the minimal orbits of the coadjoint representation
of $G_2$,
\be
X=\frac{G_2}{SU(2)\times U(1)},\quad P_X=P_5(t).
\ee

\item[(vi)] For the octonionic ``flag-like" coadjoint orbit of
$G_2$:
\be
X=\frac{G_2}{U(1)\times U(1)},\quad P_X=\frac{(1-t^4)\,
(1-t^{12})}{(1-t^2)\,(1-t^2)}=P_1(t)\,P_5(t).
\ee

\item[(vii)] For the complex Grassmann manifolds $\mathbb C
G_{m,n}$
\be
X=\mathbb C G_{m,n}=\frac{SU(m+n)}{SU(m)\times SU(n)\times
U(1)},
\ee
\be
P_X=\frac{(1-t^4)\ldots (1-t^{2(m+n)})}{(1-t^2)\left[
\left(1-t^4\right)  \ldots (1-t^{2m})\right] \left[ (1-t^4)\ldots
(1-t^{2n})\right] }.
\ee
For example, for the lowest dimensional complex Grassmann
manifold, $\mathbb C G_{2,2}$, we have
\be
X=\mathbb C G_{2,2}=\frac{SU(4)}{SU(2)\times SU(2)\times U(1)},
\ee
\be
P_X=\frac{(\,1-t^6\,)\,(\,1-t^8\,)}{(\,1-t^2\,)\,(\,1-t^4\,)}=
(\,1+t^4\,)(\,1+t^2+t^4\,)=1+t^2+ 2t^4+t^6+t^8.
\ee

\item[(viii)]  For the octonionic Cayley plane,
\be
X=\frac{F_4}{C_3\times SO(2)},\quad
P_X=\frac{(\,1-t^{16}\,)\,(\,1-t^{24}\,)} {(\,1-t^2\,)\,(\,1-t^8\,)},\ee
\be P_X=(1+t^8)(1+t^2+t^4+\cdots +t^{22})=\ee
\[ 1+t^2+t^4+\cdots +2t^8+2t^{10}+\cdots +2t^{22}+t^{24}+
\cdots +t^{30}. \]

\end{description}

\setcounter{equation}{0}
\renewcommand{\theequation}{B.\arabic{equation}}

\section*{\secfont APPENDIX B. COHERENT STATES}
\setcounter{equation}{0}

As discussed in Section 2, we consider here classical Hamiltonian
systems which  correspond to  quantum systems of a special type
for which the quantum properties are  expressed simply in terms
of classical ones.

Let $(X,\omega )$ be a compact simply-connected symplectic
manifold on which  the semi-simple compact Lie group $G$ act
transitively.

As it was shown by A. Borel \mycite{19}{Bo1}, this class of manifolds
coincides with  the class of orbits of a coadjoint or (what is
equivalent) adjoint  representation of the compact semi-simple
Lie group $G$. These manifolds   are K\"ahler homogeneous
manifolds, and have even dimension. This means that they admit a
Hermitian $G$-invariant metric, as given in
(\ref{metric}), whose imaginary part is a closed two-form given in
(\ref{twoform}). Both are determined by a single function
$F(z,{\overline z})$, called the potential of the  K\"ahler
metric, which may be found from the Gauss decomposition of the
group $G$.

The $G$-invariant Hermitian metric (and the $G$-invariant
symplectic structure) on the orbits of coadjoint actions is not
uniquely determined.  The most general ones are a linear
combination of a number $r$  of basic metrics or symplectic
forms, the number $r$ being equal to the rank of the manifold.

Let us recall now the construction of unitary irreducible
representations of  simple compact Lie groups $G$ of rank $r$.
Such representation is  characterized by an $r$-dimensional
vector $\lambda =(\lambda _1,\ldots ,
\lambda _r)$ --- the so-called highest weight: $T(g)=T^\lambda
(g)$, where
$\lambda =\sum \lambda _jw_j$, $w_j$ are the fundamental weights
and
$\lambda _j$ are non-negative integers.

Correspondingly, in the representation space ${\cal H}^\lambda $,
there exists a vector (the highest vector) $|\lambda \rangle $
satisfying the  conditions
\be \hat E_\alpha \,\vert \lambda \rangle =0,\quad \alpha \in
R_+,\quad
\hat H_j\,\vert \lambda \rangle =\lambda _j\,\vert \lambda
\rangle , \ee where $\hat E_\alpha $ and $\hat H_j$ are operators
in ${\cal H}^\lambda $ which represent the Chevalley basis for
$\cal G{}^C$.

In the space ${\cal H}^\lambda $, there exists a basis $\{ \vert
\mu  \rangle \}$, where $\mu $ is a weight vector, i.e., an
eigenvector  of all operators $H_j$: \be H_j\,\vert \mu \rangle
=\mu _j\,\vert \mu \rangle . \ee A general representation
$T^\lambda (g)$ characterized by the highest weight
$\lambda =(\lambda _1,\ldots ,\lambda _r)$ corresponds to a fiber
bundle  over $X = G/H = G^{\mathbb C}/B_+ = X_-$, with the circle
as a fiber, with  connection and curvature forms:
\be
\theta^\lambda = \frac{1}{2i}    \left( \frac{\partial F^\lambda
}{\partial
z_\mu }\,dz_\mu -\,
\frac{\partial F^\lambda }{\partial \overline z_\mu }\,d\overline z_\mu
\right) ,\qquad
\omega^\lambda  = \frac{1}{2i} \frac{\partial ^2F^\lambda }{\partial z_\mu
\,\partial
\overline z_\nu }
\,dz_\mu \land \partial\overline z_\nu =d\theta^\lambda,
\label{ConCur}
\ee
where $F=\sum _l\,\lambda _l\,F^l,\quad l=1, 2,\ldots , r.$ The
representation $T^\lambda (g)$ with the highest weight $\lambda $
may be  realized in the space of polynomials ${\cal F}^\lambda $
over $X_-$.  Namely,
\be
T^\lambda (g)\,f(z)=\alpha _\lambda (z,g)\,f(z_g),
\ee
where the quantities $\alpha _\lambda (z,g)$ and $z_g$ may be
found from the Gaussian  decomposition
\begin{eqnarray}
zg &=& \zeta _1\,h_1\,z_1, \\
z_g &=& z_1,\quad \alpha _\lambda (z,g)=\delta _1^{\lambda _1}\ldots
\delta _r^{\lambda _r}.
\end{eqnarray}
The invariant scalar product ${\cal F}^\lambda $ is introduced by
the formulas
\be
(f_1,f_2)=\,d_\lambda \int {\overline f}_1(z)\,f_2(z)\,d\mu
_\lambda (z),
\ee
where $d_\lambda $ is the dimension of the representation
$T^\lambda $.  In this case we have
\be
T^\lambda (g)\,f(z)=\exp \, [i\,S^\lambda (z,g)]\,f(z_g),
\ee
where
\be
\ \  S^\lambda (z,g) = \int _0^z (\theta^\lambda  -g_*\cdot
\theta^\lambda
)+S^\lambda (0,g),
\ee
and the K\"ahler potential is:
\begin{eqnarray}
F^\lambda = \sum
\lambda _l\,F_l^\lambda (z,{\overline z})=-\,\mbox{ln}\langle \lambda
\vert \, T^\lambda(zz^+)\,\vert \lambda \rangle,
\end{eqnarray}
which determines after (\ref{ConCur}) the connection
$\theta^\lambda$  and curvature $\omega^\lambda$ forms in the
fiber bundle with base $X$, a circle as a fiber, and related to
the representation  $T^\lambda (g)$.

A similar construction works also for degenerate representations
for which  the highest weight $\lambda $ is singular, i.e.
$(\lambda ,\alpha )=0$ for  one or several roots $\alpha $. Then
the {\em isotropy} subgroup $\tilde B$   of a vector $\vert \psi
_0\rangle $ is one of the so-called parabolic  subgroups. This
means that ${\tilde B}$ contains the Borel subgroup $B$, i.e.
the maximal solvable subgroup. The coset space $X=G^C/{\tilde B}$
is the  degenerate orbit of the coadjoint representation, but
this space is still  the homogeneous K\"ahler manifold
\mycite{19}{Bo1}. Hence the construction  considered above is valid
completely also in this case.

Following  \mycite{6,7,8}{Pe1, Pe2, Pe3}, let us now construct the
coherent
state (hereafter CS) systems for an arbitrary compact Lie group.

To this aim one has to take an initial vector $\vert 0\rangle $
in the space ${\cal H}^\lambda $. Note first of all that the
isotropy subgroup  $H_\mu $ for any state $\vert \mu \rangle $
corresponding to weight vector  $\mu $ contains the Cartan
subgroup $H=U(1)\times \cdots \times U(1)=T^r$, where $r$ is the
number of $U(1)$ factors entering in $H$, and is called the rank
of group $G$. For generic weight vectors subgroup $H_\mu
$ coincides with $H$.

In general, the isotropy subgroup for a linear combination of
weight vector  is a subgroup of the Cartan subgroup. Therefore it
is convenient to choose  a weight vector $\vert \mu \rangle $ as
an initial element of the CS system.  In the general case, the
isotropy subgroup $H_\mu $ is isomorphic to the  Cartan subgroup
$H$, and a CS is characterized by a point of $X=G/H$.

For the degenerate representation, where the highest weight
$\lambda $ is  orthogonal to some root $\alpha :(\lambda ,\alpha
)=0$, the isotropy subgroup
$H_\mu $ may be larger than $T^r$ for some state vector $\vert
\mu \rangle $.  Then any CS $\vert x\rangle $ is characterized by
a point of a degenerate orbit  of the adjoint representation.
Indeed, in all cases,
\be
H_j^\prime \,\vert x\rangle =\left[ T(g)\,H_j\,T^{-1}(g)\right]
\vert x\rangle =\mu _j\,\vert x\rangle ,\quad \vert x\rangle
=T(g)\, \vert \mu \rangle .
\ee
Therefore if we take a state vector $\vert \mu \rangle $ as the
initial  vector $\vert 0\rangle $, then the coherent state $\vert
x\rangle $ is  characterized by a point of an orbit of adjoint
representation, and the orbit  may be degenerate.

Now suppose that $T^\lambda (g)$ is a non-degenerate
representation of the  compact Lie group $G$ with the highest
weight $\lambda $, i.e., $(\lambda ,  \alpha )\neq 0$ for any
$\alpha \in R$. We take the vector with the  lowest weight $\vert
-\lambda \rangle $ as the initial vector $\vert 0 \rangle $ for
the CS system. Let us consider the action on this state  of
operators $H_j$, $E_\alpha $ and $E_{-\alpha }\,\,(\alpha \in
R_+)$  representing the Lie algebra ${\cal G}^C$. One can see
that subalgebra  ${\cal B}_-=\{ H_j, E_{-\alpha }\},\,\,\alpha \in
R_+$ is the isotropy  subalgebra for the vector $\vert \lambda
\rangle $. The corresponding group
$B_-$ is a subgroup of $G^C$.

Taking the lowest weight vector $\vert \lambda \rangle $ as
$\vert 0 \rangle $, applying operators $T^\lambda (g)$ and using
the Gaussian  decomposition $g=\zeta hz$, with $\zeta \in Z_+$, we
obtain the CS system
\be
\vert \zeta \rangle =N\,T^\lambda (\zeta )\,\vert 0\rangle =
N\,\exp \left( \sum _{\alpha \in R_+}\zeta _\alpha
\,E_\alpha \right) \vert 0\rangle,
\qquad N=\langle
0\vert\,T^\lambda (g)\,\vert 0\rangle ,\nonumber
\ee
or in another form,
\be
\vert \zeta \rangle =D(\xi )\,\vert 0\rangle ,\quad D(\xi )=\exp
\left[ \sum \left( \xi _\alpha \,E_\alpha -\overline {\xi }_\alpha \,E_
{-\alpha }\right) \right] .
\ee

Note that the unitary operators $D(\xi )$ do not form a group but
their  multiplication law is
\be
D(\xi _1)\,D(\xi _2)=D(\xi _3)\,\exp \left( i\sum _j\varphi
_j\,H_j \right) .
\ee
Note also that these CS are eigenstates of operators
\be
T(g)\,H_j\,T^{-1}(g)=\tilde H_j,\quad \tilde H_j\,\vert x\rangle =
-\,\lambda _j\,\vert x\rangle .
\ee
The last equations determine the CS up to a phase factor
$\exp(i\alpha )$.  The constructed CS system has all properties
of a general CS system. Some  of the most important ones are
noted below.

\begin{enumerate}
\item Operators $T^\lambda (g)$ transform one CS into another,
\be T^\lambda (g)\,\vert x\rangle =\exp (i\,\phi _\lambda (x,g))\,
\vert x_{g}\rangle , \ee where $\phi _\lambda (x,g)$ is a phase
shift.

\item  CS are not mutually orthogonal. The scalar product is
\[ \langle \zeta _1\vert \zeta _2\rangle =N_1\,N_2\,\langle
0\vert\, T^+(\zeta _1)\,T(\zeta _2)\,\vert 0\rangle
=N_1\,N_2\,\langle 0\vert \, T(\zeta _1^+\,\zeta _2)\,\vert
0\rangle \]
\be =K_\lambda (\zeta _1^+\,\zeta _2)\,\left[ K_\lambda (\zeta
_1^+\zeta _1)\, K_\lambda (\zeta _2^+\,\zeta _2)\right] ^{-1/2},
\ee
where
\[ K_\lambda (\zeta _1^+\,\zeta _2)=\Delta _1^{\lambda _1}(\zeta
_1^+\,
\zeta _2)\ldots \Delta _r^{\lambda r}(\zeta _1^+\zeta _2) \]
and quantities $\Delta _j$ may be found from the Gaussian
decomposition. For  the group $G=SU(n)$, $G^C=SL(n, {\mathbb
C})$, the quantity $\Delta _j$ is the  lower angular minor of
order $j$ of the matrix $\zeta _1^+\,\zeta _2$.
\end{enumerate}

%% \begin{thebibliography}{9999999}

\section*{\secfont REFERENCES}

\noindent \mycite{1}{} % \bibitem[Be 1984]{Be}
Berry M.,
Proc. Roy. Soc. London {\bf A392}, 45--57 (1984)

\noindent \mycite{2}{} % \bibitem[Si 1983]{Simon}
Simon B.,  Phys. Rev Lett. {\bf 51}, 2167 (1983)

\noindent \mycite{3}{} % \bibitem{[AA 1987]{AA}}
Aharonov Y. and Anandan,  J., Phys. Rev. Lett.
{\bf 58}, 1593 (1987)

\noindent \mycite{4}{} % \bibitem[BBK 1991]{BoBoKe}
B\"ohm A., Boya L.J., Kendrick B.,
Phys. Rev. A, {\bf 43}, 1206  (1991)

\noindent \mycite{5}{} % \bibitem[GPP 1989]{GPP}
{\em Geometric Phases in Physics},
Ed. by A. Shapere  and F. Wilczek, World Scientific: Singapore  (1989)

\noindent \mycite{6}{} % \bibitem[Pe 1972]{Pe1}
Perelomov A.M.,  {\em Coherent states for
arbitrary Lie group}, Commun. Math. Phys. {\bf 26}, No.3, 222--236  (1972)

\noindent \mycite{7}{} % \bibitem[Pe 1979]{Pe2}
Perelomov A.M.,
{\em Description of generalized  coherent states
closest to classical states},
Sov. J. Nucl. Phys. {\bf 29},  No.6, 867--871 (1979)

\noindent \mycite{8}{} % \bibitem[Pe 1986]{Pe3}
Perelomov A.M.,  {\em Generalized Coherent
States and  Their Applications}, Springer--Verlag: New-York (1986)

\noindent \mycite{9}{} % \bibitem[Bo 1955]{Bo2}
Borel A.,
Bull. Am. Math. Soc. {\bf 61}, 397--432 (1955)

\noindent \mycite{10}{} % \bibitem[CdW 1982]{CdWitt}
Choquet Bruhat, Y., de Witt Morette, C., {\em Analysis,
Manifolds and Physics}, Part I, North Holland  (1982)

\noindent \mycite{11}{} % \bibitem[BCG 1994]{BoCaGra}
Boya L.J., Cari\~nena J.F.,
Gracia-Bondia J.M., Phys. Lett A, {\bf 161}, 30-34 (1991)

\noindent \mycite{12}{} % \bibitem[Ch 1967]{Chern}
Chern, S. S.,
{\em  Complex Manifolds without Potential Theory},
 D. van Nostrand Co., Princeton, N.J. (1967)

\noindent \mycite{13}{} % \bibitem[Na 1990]{Nak}
Nakahara M., {\em  Geometry, Topology and
Physics}, Adam Hilger,  Bristol.  (1990)

\noindent \mycite{14}{} % \bibitem[Be 1999]{Be1}
Berceanu S.,
 {\em Coherent states, phases and symplectic area of geodesic
triangles}, math. DG/9903190.

\noindent \mycite{15}{} % \bibitem[FNOS 1992]{FeNiOlSa}
Fern\'andez-C D.J., Nieto L. M.,
Olmo M. A. and  Santander M., J. Phys. A, {\bf 25}, 5151 (1992)

\noindent \mycite{16}{} % \bibitem[FOS 1992]{FeOlSa}
Fern\'andez-C, D.J., Olmo M. A.
and Santander M.,  J. Phys. A, {\bf 25}, 6409 (1992)

\noindent \mycite{17}{} % \bibitem[Ar 1978]{Ar}
Arnold V.I., {\em Mathematical Methods of
Classical  Mechanics}, Springer-Verlag. New York (1978)

\noindent \mycite{18}{} % \bibitem[Ki 1976]{Ki}
Kirillov A.A.,
{\em Elements of the Theory of  Representations},
Springer--Verlag: Berlin  (1976)

\noindent \mycite{19}{} % \bibitem[Bo 1954]{Bo1}
Borel A.,
Proc. Nat. Acad. Sci. USA {\bf 40}, 1147--1151  (1954)

\noindent \mycite{20}{} % \bibitem[Go 1954]{Go}
Goto M.,
Am. J. Math. {\bf 76}, 811--818 (1954)

\noindent \mycite{21}{} % \bibitem[AP 1986]{AP}
Alekseevsky D.V. and Perelomov A.M.,
{\em Invariant  K\"ahler--Einstein metrics on
compact homogeneous spaces},
Funct. Anal. Appl. {\bf 20}, 171--182 (1987)

\noindent \mycite{22}{} % \bibitem[He 1978]{He}
Helgason S., {\em Differential Geometry,
 Lie Groups, and Symmetric Spaces}, Academic Press: NY  (1978)

\noindent \mycite{23}{} % \bibitem[Wo 1978]{Wo}
Wolf J.,  {\em Representations associated to minimal co-adjoint
orbits}: in {\em Lecture Notes in Math.} {\bf 676},  329--349
(1978)

\noindent \mycite{24}{} % \bibitem[Hu 1963]{Hu}
Hua L.-K., {\em Harmonic Analysis of
   Functions of Several Variables in Classical Domains}, AMS,
Providence, RI  (1963)

\noindent \mycite{25}{} % \bibitem[Hi 1966]{Hirz}
Hirzebruch U., Math. Zs., {\bf 115}, 387-390, (1966)

\noindent \mycite{26}{} % \bibitem[Ch 1955]{Ch}
Chevalley C.,
Am. J. Math. {\bf 77}, 778--782  (1955)

%% \end{thebibliography}
\end{document}